\documentclass[preprint]{aastex61}

\usepackage{amsmath}
\usepackage{mathtools}
\usepackage{graphicx}
\usepackage[utf8]{inputenc}
\shorttitle{BRITE photometry of $\beta$~Lyrae}
\shortauthors{Rucinski et al.}

\begin{document}

\title{
Light-curve instabilities of $\beta$~Lyrae observed by the BRITE satellites
}

\author{Slavek M.\ Rucinski} 
\affiliation{Department of Astronomy and Astrophysics, 
University of Toronto,\\ \qquad 50 St.~George St., Toronto, Ontario, M5S~3H4, Canada}
%\email{rucinski@astro.utoronto.ca}
\author{Andrzej Pigulski}
\affiliation{Instytut Astronomiczny, Uniwersytet Wroc{\l}awski, 
Kopernika~11, 51-622 Wroc{\l}aw, Poland}
%\email{pigulski@astro.uni.wroc.pl}
\author{Adam Popowicz}
\affiliation{Instytut Automatyki, Wydzia{\l} Automatyki Elektroniki 
i Informatyki, Politechnika \'{S}laska,\\ \qquad  Akademicka~16, 44-100 Gliwice, Poland}
%\email{Adam.Popowicz@polsl.pl}
\author{Rainer Kuschnig}
\affiliation{Institute of Communication Networks and Satellite Communications, 
   Graz University of Technology,\\ \qquad Inffeldgasse~12, 8010 Graz, Austria}
%\email{rainer.kuschnig@tugraz.at}
\author{Szymon~Koz{\l}owski}
\affiliation{Warsaw University Observatory, Al.~Ujazdowskie~4, 00-478 Warszawa, Poland}
%\email{simkoz@astrouw.edu.pl}
\author{Anthony F. J. Moffat} 
\affiliation{D\'{e}partement de physique, Universit\'{e} de Montr\'{e}al, C.P. 6128, 
   Succursale Centre-Ville,\\  \qquad Montr\'{e}al, Qu\'{e}bec, H3C 3J7, Canada}   
%\email{moffat@astro.umontreal.ca}
\author{Kre\v{s}imir Pavlovski}
\affiliation{Department of Physics, Faculty of Science, University of Zagreb, 
Bijeni\v{c}ka cesta 32, 10000 Zagreb, Croatia}
%\email{pavlovski@phy.hr}
\author{Gerald Handler}
\affiliation{Nicolaus Copernicus Astronomical Center, Bartycka~18, 00-716 Warszawa, Poland}
%\email{gerald@camk.edu.pl}
\author{H.\ Pablo}
\affiliation{American Association of Variable Star Observers, 49~Bay State Road, 
Cambridge, MA~02138, USA}
%\email{hpablo@aavso.org}
\author{G.\  A.\  Wade}
\affiliation{Department of Physics and Space Science, 
Royal Military College of Canada,\\  \qquad PO Box~17000 Kingston, Ontario, K7K~7B4, Canada}
%\email{gwade@queensu.ca}
\author{Werner W.\ Weiss}
\affiliation{Institut f\"{u}r Astrophysik,Universit\"{a}t Wien, T\"{u}rkenschanzstrasse~17, 1180~Wien, Austria}
%\affiliation{University of Vienna, Institute for Astrophysics, Tuerkenschanzstrasse 17, Vienna, Austria}
%\email{werner.weiss@univie.ac.at}
\author{Konstanze Zwintz}
\affiliation{Institut f\"{u}r Astro- und Teilchenphysik, Universit\"{a}t Innsbruck, Technikerstrasse~25, 
A-6020 Innsbruck, Austria}

%\email{Konstanze.Zwintz@uibk.ac.at}

%\author{the BRITE Team}

\correspondingauthor{Slavek M. Rucinski}         % as per the AAS template
\email{rucinski@astro.utoronto.ca}

%\correspondingauthor{Slavek M. Rucinski}      %  the ApJ emulation does not accept this
%\email{rucinski@astro.utoronto.ca}

\begin{abstract}
Photometric instabilities of $\beta$ Lyr were observed in 2016 by two red-filter BRITE
 satellites over more than 10 revolutions of the binary, with $\sim$100-minute sampling. 
 Analysis of the time series shows that flares or fading events take place typically 
 3 to 5 times per binary orbit. The amplitudes of the disturbances (relative to 
 the mean light curve, in units of the maximum out-of-eclipse light-flux, f.u.) are 
 characterized by a Gaussian distribution with $\sigma=0.0130\pm0.0004$ f.u. 
 Most of the disturbances appear to be random, with a tendency to remain for one
  or a few orbital revolutions, sometimes changing from brightening to fading or 
  the reverse. Phases just preceding the center of the deeper eclipse showed 
  the most scatter while phases around secondary eclipse were the quietest. 
  This implies that the invisible companion is the most likely source of the 
  instabilities. Wavelet transform analysis showed domination of the variability 
  scales at phase intervals $0.05-0.3$ (0.65--4 d), with the shorter (longer) scales 
  dominating in numbers (variability power) in this range.  The series can be well 
  described as a stochastic Gaussian process with the signal at short timescales 
  showing a slightly stronger correlation than red noise. The signal de-correlation timescale 
$\tau=(0.068\pm0.018)$ in phase or $(0.88\pm0.23)$~d appears to follow the same
 dependence on the accretor mass as that observed for AGN and QSO masses 5--9 
 orders of magnitude larger than the $\beta$~Lyr torus-hidden component.
 \end{abstract}

\keywords{stars: individual ($\beta$~Lyr) - binaries: eclipsing 
- binaries: close - binaries: photometric - techniques: photometric}

\section{Introduction} 
\label{sec:intro}

$\beta$~Lyr (HD~174638, HR~7106) is a frequently observed, bright 
 ($V_{\rm max}=3.4$ mag, $B-V=0.0$ mag) eclipsing binary. Hundreds of papers have 
been published about this complex system, which
consists of a B6-8~II bright giant with a mass of about $3 M_\odot$ 
and an invisible, much more massive companion ($\simeq 13 M_\odot$)
which occults the blue bright-giant every $P = 12.915$ days producing
primary (deeper) eclipses.
The B-type bright giant loses mass to the more massive object at a rate which induces a 
fast period change, $dP/dt$ of 19 seconds per year; this period change has been 
followed and verified for almost a quarter of a millennium\footnote{The beautiful parabolic 
shape of the continuously updated, {\it observed minus calculated\/}  
times-of-minima ($O-C$) diagram can be appreciated by inspection of    
on-line material in {\it http://www.as.up.krakow.pl/minicalc/LYRBETA.HTM}
\citep{Krein2004}.}. 
The binary shares many features with the W~Serpentis
binaries which are characterized by the presence of highly excited ultraviolet
spectral lines most likely energized by the dynamic mass-transfer between the
components \citep{Guin1989,Plav1989}. 
The accretion phenomena related to the mass transfer from 
the visible component onto the more massive component lead to processes resulting in
complex spectral-line variability, with the presence of strong emission lines 
\citep{Harm1996, Ak2007}, of X-ray emission \citep{Ign2008} 
and of variable spectral polarization \citep{Lom2012}. The physical
nature of the more massive component of $\beta$~Lyr remains a mystery, but it is
normally assumed that it is a donut-shaped object with the outer regions
completely obscuring the view of a central star expected to be roughly of B2 spectral type. 
The interferometric observations by the CHARA system confirm
this general picture \citep{Zhao2008}. 
For further details, consult the review of the physical properties of $\beta$~Lyr 
by \citet{Harm2002} which is a useful summary of  what is known and 
what remains to be learned about the binary system.

Accretion phenomena in $\beta$~Lyr lead to light-curve instabilities. A dedicated 
international, multi-observatory campaign in 1959 \citep{LL1969a,LL1969b} 
led to detection of instabilities as large as 0.1 mag. However, observational errors during 
this campaign were typically about 0.01 mag and temporal characteristics of the instabilities 
could not be firmly established. Because of the need for unifying data from many observatories, 
there remained a lingering possibility that the apparent light-curve
instabilities could be at least partly explained by photometric calibration or
filter-mismatch problems and by the presence of emission lines coupled to 
differences in photometric filters. All these problems were aggravated even further
by the moderately long orbital period and by the diurnal interruptions. 
Although the 1959 photometric campaign left no doubt that the instabilities are 
 present, their frequency and size remained poorly understood. 
In this situation, attempts to model the light curves 
\citep{Wil1974,vHW1995,Linn1998,Menn2013} 
were always confronted with a necessity to use orbital-phase averages 
with the assumption that the photometric instabilities are sufficiently
random to permit such an averaging to obtain reliable mean light-curve values. 

This paper sets as a goal characterization of the light-curve instabilities 
observed in $\beta$~Lyr by the BRITE  Constellation
\citep{Weiss2014,Pablo2016,Pop2017} during  5 months in 2016
using the temporal sampling of the orbital periods of the satellites, i.e.\ 
98 -- 100 minutes. 
Section~\ref{sec:obs} describes the observations while Section~\ref{sec:lc}
presents a mean light curve used in this paper as an auxiliary tool to derive
photometric deviations from it. 
Section~\ref{sec:instab} discusses analyses of the light-curve instabilities treated 
as a time series using traditional tools such as the Fourier transform or
autocorrelation function, and adding newer methods such as wavelets
and recently developed methods to characterize variability of QSO and AGN objects.
The concluding Section~\ref{sec:concl} summarizes the results.

\section{BRITE observations}
\label{sec:obs}

\subsection{General description}
\label{sec:gen-obs}

The observational material obtained during the 2016 visibility season 
of $\beta$~Lyr consists mainly of photometric observations from 
two satellites of the BRITE  Constellation equipped with red filters, 
``BRITE-Toronto'' (BTr) and ``Uni-BRITE'' (UBr). 
The star was  observed also by the blue-filter satellite, 
``BRITE-Lem'' (BLb), but this satellite suffered
from stabilization problems so that the observations covered 
less than one orbital period of the binary at the end of 
the BTr observations; as a result, the BLb data are not used in this paper.

The BRITE observations started on 4 May and ended on 3 October 2016
and thus lasted 152 days. The exposures were taken at 20-second 
intervals with the duration of 1 second for the UBr satellite, which is a
typical choice for bright stars, preventing saturation of the CCD pixels.
Unfortunately, UBr experienced stabilization problems 
for about a half of the  $\beta$~Lyr run, so that most of the data were collected 
by the BTr satellite using 3-second exposures, which were more appropriate 
for $\beta$~Lyr. BTr was very well stabilized and provided excellent data
in terms of their quality and quantity.

% --------------------------------------- fig1--------------------------------------------------------
\begin{figure}[ht]
\begin{center}
\includegraphics[width=0.48\textwidth,angle=90]{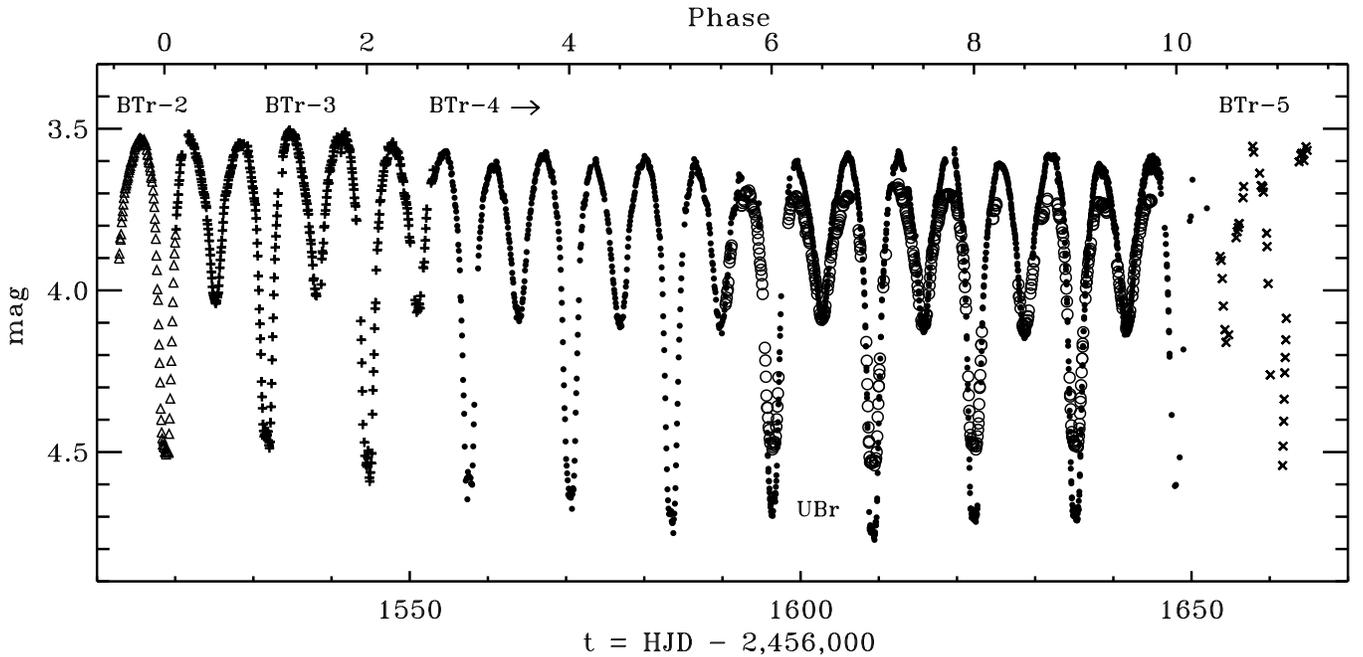}    
\caption{
BRITE observations of $\beta$~Lyrae in 2016 versus
time in $t = HJD- 2\,456\,000$. The satellite-orbit
averages (in magnitudes) shown in the figure resulted from standard pipeline 
and de-correlation processing. 
The orbital phases of the binary are given along the upper horizontal 
axis of the figure; they are counted continuously from the assumed zero epoch,
as described in Section~\ref{sec:ph}.
The symbols and labels relate to different satellite ``setups''
as described in the text (BTr-2: triangles; BTr-3: crosses; BTr-4: filled circles;
BTr-5: slanted crosses; UBr: open circles). 
While the magnitude system is arbitrary, the {\it amplitude\/} of the BTr 
variability is larger compared with that observed by UBr; {\bf  note the light-curve 
shape defined by the small, filled circles and by the overlapping larger, open circles.} 
This unexpected result led to the discovery of an instrumental 
problem which is described in Section~\ref{sec:instr}.
All available red-filter observations are shown here including those which
were not used in the detailed analysis limited to the BTr-3, BTr-4 and
UBr setups.  The blue-filter BLb observations, which are not shown, 
took place at the end of the BTr-4 run, at $1637 < t < 1647$ 
and covered about 0.7 of the binary orbital period.
}
\label{fig_mags}
\end{center}
\end{figure}
% -----------------------------------------------------------------------------------------------------

The distribution of the BRITE magnitudes versus time 
%($t = HJD - 2\,456\,000$)
is shown in Figure~\ref{fig_mags}. 
Since the BRITE data are used mainly for studies of stellar 
variability, the zero point of the magnitude scale is arbitrary and 
is adjusted for each of the ``setups'' marked by labels in the figure. 
A ``setup'' is a set of positioning instructions for the satellite 
and for the CCD windowing system, as described in full in the
Appendix to \citet{Pop2017}. 
Occasional changes of setups result from addition or removal of stars from
the observed field without changes of the remaining star positions, while
some changes are necessary because of satellite stability
issues or Earth-shine scattering problems. 
The concept of the ``setup'' is important, as it divides the data
into series which may show small magnitude shifts. 
The division into setups continues through the initial raw-data processing
so that each setup can be identified as leading to a separate time 
series. Several setups were inadequate for the current investigation:
The first BTr setup could not be used because of
imperfections in the initial positioning, the BTr-2 was shorter
than one orbital period of $\beta$~Lyr, while the data for the
last setup BTr-5 showed strong signatures of being affected 
by  scattered sunlight at the end of the season. In the end, 
this paper utilizes the data for the setups BTr-3, BTr-4 and the
combined data for the setups UBr-1 and UBr-2 (from now on, UBr).
The ranges in time and the numbers of observations in each setup 
are listed in Table~\ref{tab:setups}. 

\placetable{tab:setups}                      % Table 1   individual setups

\subsection{Discovery of an instrumental problem}
\label{sec:instr}

The data processing followed the routine steps described in 
\citet{Pablo2016} and \citet{Pop2017} with subsequent de-correlation 
as described by \citet{Cookbook2}.
Even a cursory comparison of the raw data (Figure~\ref{fig_mags})
reveals a large difference in the amplitude of $\beta$~Lyr as measured 
by UBr and BTr, with BTr observing amplitudes larger than one magnitude, a value 
never encountered  before. It was obvious that a previously unrecognized 
problem affected the BTr observations. It went undetected because none
of the previously observed variables had such a large and well known amplitude. 
Detailed comparison of the resulting light-curve shapes 
and lack of any indications for photometric non-linearity
in the system led us to consider a linear transformation of the 
detected signal involving a loss of detector charge, somewhat
similar to a locally different CCD bias. 
Further investigation revealed shallow spots in the CCD response
where electrons were trapped by charge-transfer-inefficiency (CTI)
effects, apparently due to radiation damage. This problem was reported
before in a  note about the same BRITE observations 
 \citep{Ruc2018}. Unfortunately, there were no contemporary 
data on the spots for the time of the $\beta$~Lyr run, but from
archival data it was estimated that in April 2015 --
when an edge part of a blank field was affected by scattered sunlight
-- the depressions covered about 2.5 to 4 percent of
the BTr detector area. The new data taken in July 2017
indicate that the amount of the affected area has grown in time
as a result of progressing detector damage. 
Discovery of the problem by our observations has opened up a full
investigation which is currently ongoing. In addition to the note
by \citet{Ruc2018}, the problem has been discussed with other instrumental 
issues affecting the BRITE satellites by \citet{Pig2018} and \citet{Pop2018}. 

The problem was detected mostly thanks to the large and well-known 
amplitude of $\beta$~Lyr; it has minimally affected most of the 
previous studies concentrated on small-scale pulsational variabilities.
In fact, since this paper is devoted to deviations from the smooth light
curves, we could ignore this new CTI effect and use the otherwise
excellent BTr data without any correction. Such results would be
spoiled by a scale problem, though, and the mean light curve would be
entirely invalid. 
Fortunately, it was  possible to correct the BTr data by using 
the partially simultaneous data from the UBr satellite. In doing
this correction, we assumed that the UBr signal 
was not systematically modified in any way and that the proper BTr signal can be 
restored by a simple linear transformation relating the CCD charges 
(or  light fluxes)  $f_{\rm UBr}$ and $f_{\rm BTr}$:
$f_{\rm UBr} = a_0 + a_1 f_{\rm BTr}$. Here, $a_0$ and $a_1$ are constants 
to be determined by least-squares fits, with $a_0/a_1$ representing
the lost charge expressed relative to the maximum signal for
$\beta$~Lyr as observed by UBr ($f_{\rm UBr}$ was normalized to unity
at the binary light maxima). 
The temporal constancy of the problem is obviously an assumption, but it 
seems to be a plausible one in view of (1)~the stability of the final results
and (2)~the most likely cause of the CTI damage as due to infrequent 
local damage of the CCD lattice caused by very energetic radiation particles.  
The fluxes were determined by the standard flux-magnitude relation,
$f = 10^{-0.4 (m - m_0)}$,  
assuming the maximum light ($m_0$) magnitudes for the setups: 
$m_0 ({\rm UBr}) = 3.81$, $m_0 ({\rm BTr2}) = m_0 ({\rm BTr3}) = 3.54$, 
and $m_0 ({\rm BTr4}) = 3.59$, 
where the magnitudes are those resulting from the standard BRITE 
pipeline and de-correlation processing. 
The transformation linking the UBr and BTr-4 data was possible 
for 50.14 days (or 3.87  $\beta$~Lyr orbital periods) of simultaneous 
observations of  the two setup sequences.
Thanks to the very high number of overlapping observations of
the UBr run (16,686), the transformation relating the $f_{\rm UBr}$
and the $f_{\rm BTr}$ fluxes was very well determined: 
$a_0 = 0.20392 \pm 0.00046$, $a_1 = 0.79688 \pm 0.00063$,   
with the amount of the signal lost, in terms of the maximum
$\beta$~Lyr signal: $a_0/a_1 = 0.25589 \pm 0.00077$.
The errors here were determined by 10,000 bootstrap repeated solutions.  
No systematic temporal trend was detected in the transformation. 

\subsection{Satellite-orbit average data}
\label{sec:sat}

Observations of $\beta$~Lyr by both satellites experienced breaks due to
Earth eclipses which naturally divided the individual observations into groups
separated by  gaps when the field was either invisible due to Earth occultations
or strongly affected by the Earthshine.
The median number of individual observations in 
the groups was 40 for BTr and 37 for UBr. Thus, with 3 observations per minute,
the uninterrupted observations lasted typically 12 -- 13 minutes.
However, some groups were as long as 75 continuously acquired observations 
and some as short as 8 observations, so that the averaged data 
have different  errors: For BTr, the errors range
between 0.0005 and 0.0045 with the median error (assumed to be typical) 0.0014;  
for UBr, the errors range between 0.0004 and 0.0058
with the median error 0.0019, all expressed in flux units (f.u.) relative to  
the maximum $\beta$~Lyr flux\footnote{Individual observations were of different
quality so that the mean errors of the averages do not simply reflect 
the Poissonian number statistics.}. The better quality of the BTr data 
is  a result of the longer individual exposures of 3 sec, compared with 1 sec
for UBr.

The satellite orbital periods at the time of the $\beta$~Lyr observations were:
 0.06824 d = 98.3 min for BTr and 0.06974 d = 100.4 min for UBr. 
 Expressed in units of the  
binary-star period, the satellite-caused spacing in $\beta$~Lyr orbital phase
was 0.005273 for BTr and 0.005388 for UBr, or 190 and 186 
group averages per one orbital period of the binary, setting respective upper
limits on frequencies of detectable instabilities. 
The satellite-orbit average fluxes of $\beta$~Lyr are given in Table~\ref{tab:avg}.
The table lists the mean time, the orbital phase as in Section~\ref{sec:ph},
the average flux with its error, and the number of observations per average. 

\placetable{tab:avg}                     % Table 2   satellite orbit averages

%\newpage

\subsection{The orbital-phases of $\beta$~Lyr}
\label{sec:ph}

The orbital period of $\beta$~Lyr has been studied by many investigators.
It is now well established that the period change rate is very close to being
constant and that a parabolic $O-C$ diagram for the times of minima
very well describes the expected moments of eclipse minima. 
We used  the elements of \citet{Ak2007} to set a locally
linear system of phases for the epoch $E=3811$: 
$\phi = (t - 1518.6251)/P$, with $P=12.943296$ corrected for the $dP/dt$ change
and $t$ as defined in Section~\ref{sec:gen-obs}. 

From now on in this paper, we will use the term ``orbital phase'' or just ``phase'' as the main
independent variable versus which the physical variability of $\beta$~Lyr
is taking place. The phase is counted as a number including the integer part,
as shown along the upper horizontal axis of Figure~\ref{fig_mags}. 
Traditionally, the meaning of the term ``phase'' is 
a number confined to the interval 0 to 1; such a use 
as a fractional phase appears sparingly in this paper.
% and is then noted as a ``fractional phase''.
The phases of our observations cover the range from about zero for the (not used) setup 
BTr-2 data to about eleven at the end of the (also not used) setup BTr-5. 
The two BTr setups which were utilized cover the phase ranges $0.11 < \phi < 2.66$ and  
$2.66 < \phi < 10.30$ for BTr-3 and BTr-4, respectively.
This system of local phases is used in the paper as 
an independent variable in place of the time in our discussion 
of the photometric instabilities treated as a time series (Section~\ref{sec:instab}).

% --------------------------------------- fig2--------------------------------------------------------
\begin{figure}[ht]
\begin{center}
\includegraphics[width=0.76\textwidth]{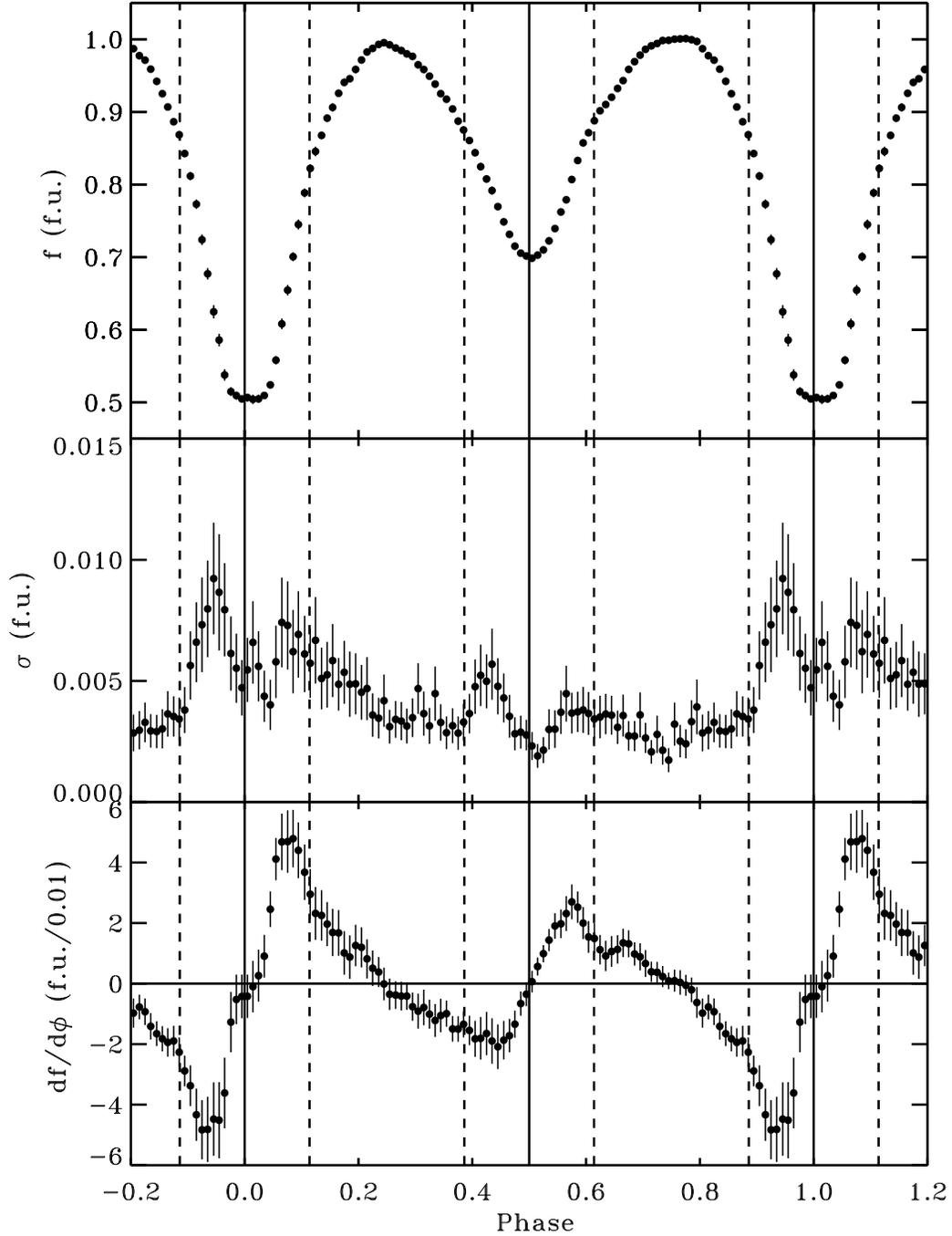}    
\caption{
The mean light curve of $\beta$~Lyrae 
obtained by averaging satellite-orbit means for the
setups BTr-3 and BTr-4 in intervals of 0.01 in  fractional orbital phase. 
The light curve is expressed in relative flux units (f.u.) with the maximum light assumed 
as unity. The error bars are comparable to or smaller than the symbols used for
the light curve (the upper panel). The errors with their uncertainties,
estimated from the number of points per phase interval, are shown separately
in the middle panel. The lowest panel gives the light-curve derivative in f.u.\  
per 0.01 phase interval {\bf with uncertainties estimated from the errors of 
adjacent flux values}. The vertical lines give the 
phases of external eclipse contacts (at $\pm 0.114$ relative to the eclipse
centers), calculated using the unpublished model by \citet{Pav2018}.
}
\label{fig_LC}
\end{center}
\end{figure}
% -----------------------------------------------------------------------------------------------------

\section{The mean light curve}
\label{sec:lc}

The mean light curve for $\beta$~Lyr has been formed from the combined
observations of the BTr-3 and BTr-4 setups. The curve itself is not used 
in this work except for removal of the global eclipsing variations and thus
to determine the light-curve instabilities. The mean light curve is very 
well defined (Figure~\ref{fig_LC}); it has been obtained by 
averaging the individual, satellite-orbit, average data points
in 0.01 intervals of binary fractional orbital phase. 
Typically (median) 16 points  per phase interval contributed
to a single point, with the actual numbers ranging between 
12 and 20 for individual satellite-orbit mean points. 
The flux error per average point ($\sigma_{\rm mean} \simeq 0.0042$ f.u., 
$\sigma_{\rm median} \simeq 0.0036$ f.u.)
is dominated by the $\beta$~Lyr light-curve instabilities which we discuss
in the rest of this paper. 
The errors show an increase during the eclipse branches as expected 
when the data are averaged on slopes and consequently correlate with the
absolute value of the derivative (the lowest panel in Figure~\ref{fig_LC}). 
Judging by the size of $\sigma_{\rm median}$ and the number of observations 
per interval, typical deviations of the instabilities are expected to 
show a scatter with $\sigma_{\rm dev} \simeq 0.014$, a number 
which is confirmed by a more detailed analysis in Section~\ref{sec:instab}.
This is in fact a smaller number than was originally expected for the
combined BTr-3/BTr-4 
observations lasting as long as four months. 
Unless we observed $\beta$~Lyr in a particularly inactive time, 
this may indicate that the previous indications of large
photometric instabilities reaching 0.1 mag were partly affected 
by inconsistencies in filter-matching and by other data-gathering 
and collation steps. 

\placetable{tab:lc}                     % Table 3  light curve

The excellent definition of the BTr light curve permitted determination
of its derivative $df/d\phi$, as shown in Figure~\ref{fig_LC}.
In addition to the expected large absolute values within the 
eclipses, the derivative reveals some structure beyond the eclipses
which may indicate phenomena not accounted for by the standard 
stellar-eclipse model. 

A phase-shift apparently caused by a small asymmetry
appears to be present for the primary (deeper) minimum for the 
time-of-eclipse elements of \citet{Ak2007}.
The shift is estimated at $+0.0076 \pm 0.0002$ in phase, 
corresponding to $+0.0984 \pm 0.0026$ day or $+2.36 \pm 0.06$ hours. 
No significant shift, at the same level of uncertainty as for the primary, 
was noted for the secondary eclipse. 
The presence of the primary-eclipse phase shift 
may be related to the 283 day periodicity in the times of
the primary eclipses which was noted before
\citep{Guin1989, Krein1999, WVH1999, Harm2002} and which still
does not have an explanation (Section~\ref{sec:283}). Since the 
secondary minimum did not show any shift, 
the phase system of \citet{Ak2007} was  adopted for our
investigation without any modification.

In addition to the BTr mean light curve, an independent light curve 
was formed from the UBr observations. It is not equally well defined 
because larger gaps in the $\beta$~Lyr phase coverage 
resulted in only 92 points for the same, 0.01-wide phase interval. 
Consequently, the deviations for the UBr series, used later to fill in 
gaps in the BTr satellite data, were defined in reference to the BTr 
mean light curve. The UBr and BTr fluxes are expected to be in the
same system through the transformation described in Section~\ref{sec:instr}.

% --------------------------------------- fig3--------------------------------------------------------
\begin{figure}[ht]
%\begin{figure}
\begin{center}
\includegraphics[width=0.67\textwidth,angle=90]{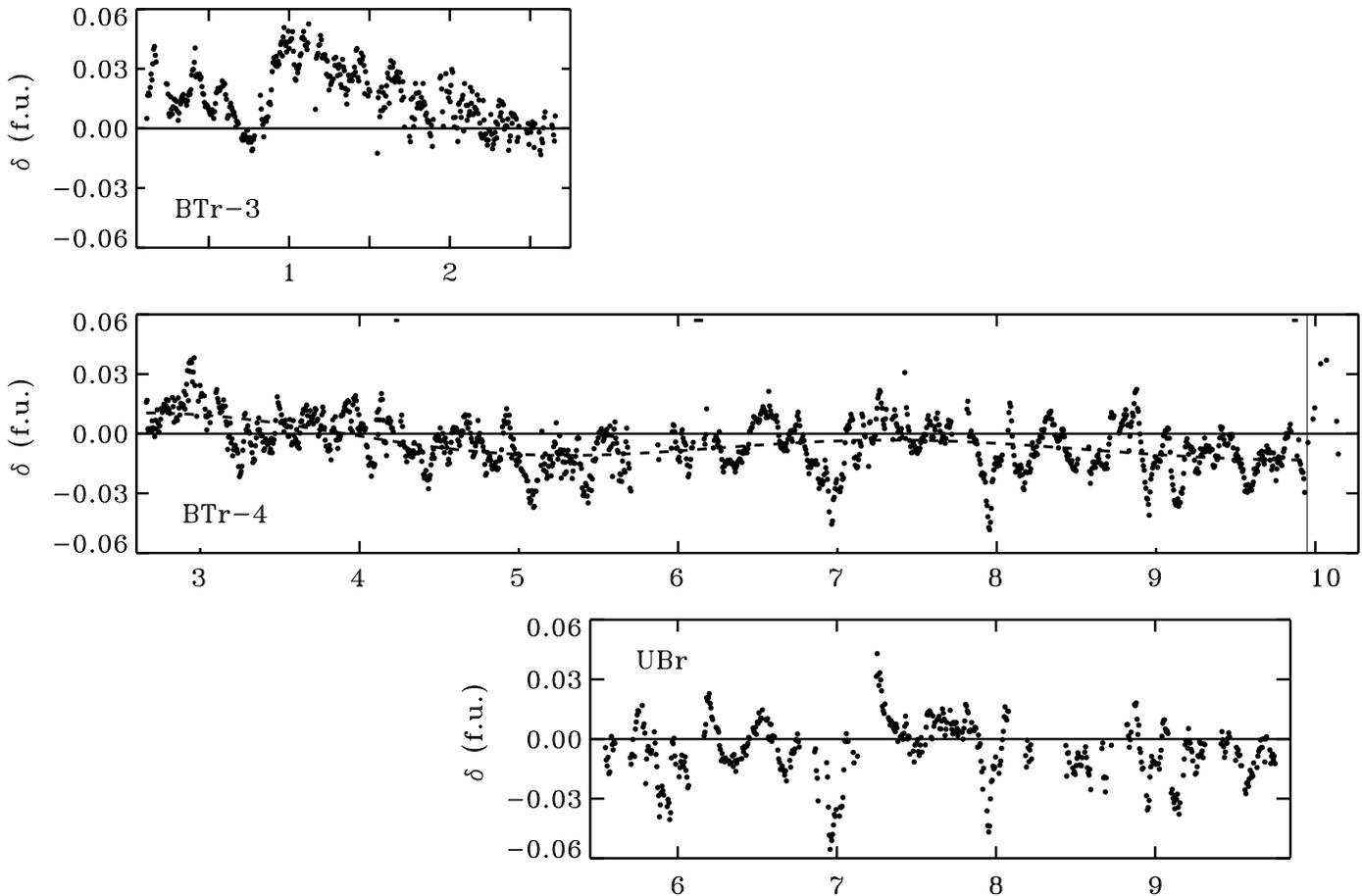}    
\caption{
The light-curve deviations $\delta$ from the mean as a function of the orbital phase
of $\beta$~Lyr. The deviations used in the analysis are for three setups:
BTr-3 (upper panel), BTr-4 (middle) and UBr (lower panel). 
The BTr-3 series precedes the BTr-4 while that for UBr
is aligned to show the same orbital phases as 
for BTr-4. The deviations $\delta$ are expressed in flux units.
% with the maximum $\beta$~Lyr light taken as unity. 
The mostly positive values of the deviations for the BTr-3 series and slightly negative ones
for the BTr-4 series result from the use of the combined BTr-3 and BTr-4 data 
for formation of the mean light curve.
The wavy, broken-line curve in the BTr-4 panel gives the low-frequency 
approximation by a 5th-order Chebyshev polynomial, as discussed 
in Section~\ref{sec:283}.
The three largest gaps in the uniform, equal-step BTr-4 coverage which could
not be filled by the UBr observations 
were interpolated over 7, 12 and 7 equal-step intervals;
they are marked  by short bars along the upper axis of the figure
{\bf (phases: 4.22--4.25,   6.10--6.16,   9.85--9.89)}.
}
\label{fig_devs}
\end{center}
\end{figure}
% -----------------------------------------------------------------------------------------------------

\section{The light-curve instabilities}
\label{sec:instab}

\subsection{The $\delta$-deviations as a time series}
\label{sec:tser}

We used deviations of the individual satellite-orbit averages
from the mean light curve to characterize
the light-curve instabilities.  The deviations, $\delta$, 
are shown versus $\beta$~Lyr orbital phase
in Figure~\ref{fig_devs}. The data are continuous 
for the adjacent BTr-3 and BTr-4 setups,  
while those provided by UBr overlap over a part of the BTr-4
series. For the BTr observations, the temporal (satellite orbit) sampling 
was 98.3 minutes, while for UBr the sampling was 100.4 minutes, with 
average deviations from uniform sampling  
of  $\pm 2.4$ minutes for BTr and $\pm 1.5$ minutes for UBr. 
The BTr-3 and BTr-4 series are of different lengths,
covering together the $\beta$~Lyr phases for 10 
orbital cycles ($0.11 < \phi < 10.30$).  
For the final application, the BTr-4 series was shortened to end at $\phi = 9.95$ 
to avoid the poorly covered observing time 
when the satellite-viewing field was getting into the Sun-illuminated part of the sky. 
Descriptions of the individual series are given in Table~\ref{tab:devs}.

\placetable{tab:devs}                     % Table 4  descriptions of equal-step series

After careful consideration of the individual time series, we decided
to use only the BTr-4 data for a study of the light-curve instabilities.
The series BTr-3 is short and shows a trend with an unexplained jump
by $\delta \simeq +0.045$ close to $\phi = 0.9$ (it is probably significant 
that the jump coincides in fractional phase with phases of increased
activity in the BTr-4 series, see below in Section~\ref{sec:freq}).
The UBr series has poor phase coverage with larger errors of 
individual data points, but it was useful to fill gaps in the BTr-4 coverage. 
It should be noted that the series
BTr-3 and BTr-4 appear to have been correctly adjusted in terms 
of the magnitude shift during the initial processing stage since
the $\delta$ deviations appear to be continuous (and fortuitously close to zero)
at the transition point at $\phi = 2.7$ (Figure~\ref{fig_devs}). 

The BTr-4 series extending into the orbital phase
interval $\phi = 2.661 - 9.948$ and sampled at satellite-orbit intervals 
with the mean $\Delta \phi =  0.0052726$ can be subject to   
%is very close to an ideal one for a 
time-series analysis. To form a perfect equal-step sequence, 
we used $\Delta \phi$ as the main unit of the equal-step grid. 
Such a series of 1383 equidistant points had gaps as 1222 satellite 
averages were actually observed. Since the UBr and BTr flux 
scales are identical through the scaling operation, 
as described in Section~\ref{sec:instr}, we were able to fill the gaps using 
the UBr data. Except for an unexplained spike observed at
$\phi \simeq 7.25$, when UBr was returning to normal operations 
after an instrumental interruption, the UBr and BTr data agree very well
(Figure~\ref{fig_devs}). Fortunately, there was no need to use the
spike phases for filling the BTr-4 series, while the UBr data filled 
very well the BTr gaps at phases $\phi \simeq 5.7$ and 7.8. 
We used UBr observations with a restriction that they could be spaced 
no more than one satellite orbit away from the missing 
interval for  interpolation into the BTr-4 series.
The addition of 77  points increased the number of observed points
to 1299 and the final coverage efficiency to 93.9 percent. 
The remaining gaps were spline-interpolated within the filled series,
typically over one to three missing points. 
The three larger gaps, one of 12 and two of 7 intervals 
do not seem to affect the shape of the series in an obvious way; they
are marked in Figure~\ref{fig_devs} as short bars along the 
upper axis of the BTr-4 panel.

\newpage

\subsection{The 283 day periodicity and its implications}
\label{sec:283}

In addition to apparently random perturbations, the light curve of $\beta$~Lyr 
is known to show a possibly periodic variation beyond that of the binary orbit, 
which has so far eluded explanation.  It was detected in the deviations from the mean 
light curve by \citet{Guin1989} who estimated their period at $275 \pm 25$ days. 
Later, using archival data, 
\citet{vHW1995} found the period of 283.4 days, while 
\citet{Harm1996} found  282.4 days, 
with similar uncertainties of about $\pm 0.1$ day. 
The maximum semi-amplitude estimated from these analyses was about 
0.03 of the mean flux. 
The period of 283 days was later confirmed by \citet{Krein1999a}. 

While the analyses by \citet{vHW1995} and by \citet{Harm1996}
showed agreement in the description of the archival material, the uncertainty in the 
period precludes forward prediction reaching the epoch of 
our observations. With the duration of 3 months, the BRITE observations 
did not last long enough to address the 283 day periodicity directly, 
however, we could look for shorter, possibly related time scales. In particular, 
\citet{Harm1996} discussed a possibility 
that the 283 day periodicity results from beating of the orbital period with  
variability at 4.7 -- 4.75 days estimated for the emission lines
emitted by a precessing jet from the accreting component.  

Our $\delta$-series does show a slow, wavy trend extending over the 
whole duration of the run (Figure~\ref{fig_devs}). However, 
since the data were affected by the problem with the missing CCD 
charges (Section~\ref{sec:instr}), our initial reaction was to treat the slow 
trend as having an instrumental explanation and to remove it. 
Such removal would obviously force the analysis to address
only the short time-scales, comparable to or shorter than the $\beta$~Lyr period. 
The slow trend was removed by utilizing the excellent approximation properties of the 
Chebyshev-polynomials: The $\delta$-series was converted into a series of such
polynomials, all their coefficients for orders $>\!5$ were set to zero 
and then the series was re-converted back to the time series. The resulting 
low-frequency wave with the range $-0.0132$ to $+0.0104$ f.u.\ 
was subtracted from the observed deviations to form a ``trend-subtracted'' series.  
%Slightly  negative values appear to dominate over the whole phase extent. 
%This is  due to the use of the combined, mean BTr-3 and BTr-4 light 
%curve which -- as we realized later --  contains a large,
%positive deviation within the phase range $1 < \phi < 2$.

\begin{figure}[ht]
\begin{center}
\includegraphics[width=0.68\textwidth]{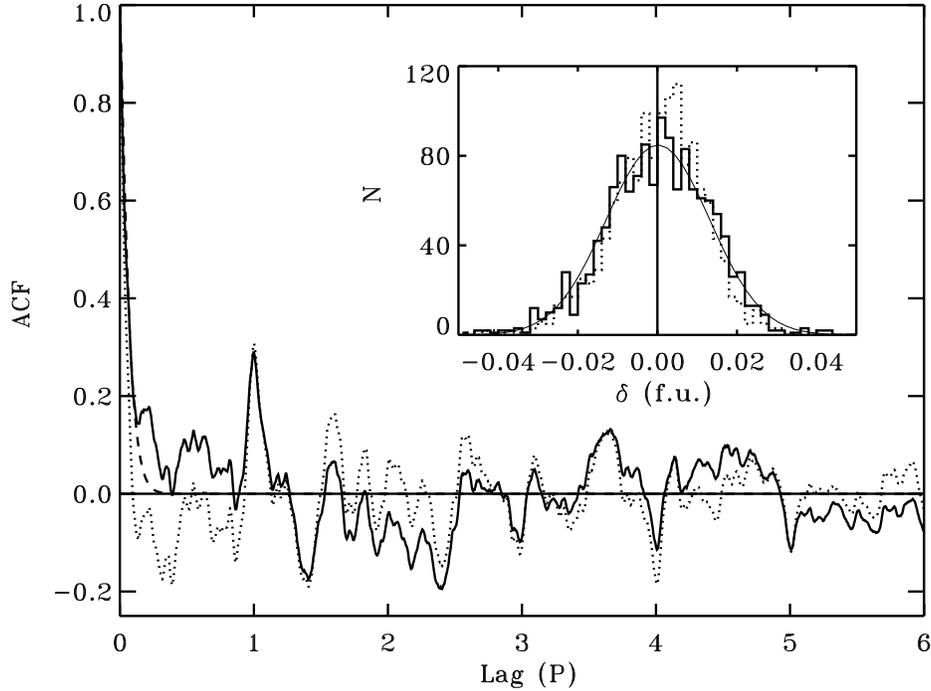}    
\caption{
The auto-correlation function (ACF) for the $\delta$ series. 
The horizontal axis gives the correlation lag
in units of the $\beta$~Lyr orbital period. The dotted line gives
the ACF for the trend-corrected series, as explained in Section~\ref{sec:283}.
The dashed line close to the origin shows the model ACF 
(Equation~\ref{eq:ACFPE}) obtained by an independent analysis of
the Structure Function (SF) for the series treated as a Damped Random Walk
process using Equation~\ref{eq:fullfit}, with $SF_\infty=0.0179$ and $\sigma_n=0.0014$; 
see Section~\ref{sec:DRW}. The insert gives the histogram of the 
versions of the $\delta$ series, with a Gaussian fit (continuous line) 
for the series as observed, without the trend correction.
}
\label{fig_acf}
\end{center}
\end{figure}

After the analysis of the trend-corrected data using tools developed for
AGN and QSO light-curve instabilities (Section~\ref{sec:DRW}) 
we realized that the uncorrected data present a more consistent picture
of a Damped Random Walk (DRW) process. 
For that reason, we carried out the analysis for both the uncorrected 
(mean-subtracted) and the trend-corrected series of deviations (Table~\ref{tab:ts})
and attempted to monitor the resulting differences. 
The deviations of both series follow a Gaussian distribution with
largest deviations reaching $\pm 0.04$ f.u.\ 
%and about $\pm 0.035$ f.u.\
for the uncorrected series.
As expected, the standard deviation for the uncorrected series is larger:
$\sigma = 0.0130 \pm 0.0004$ f.u.\ while it is 
$\sigma = 0.0109 \pm 0.0003$ f.u.\ for the trend-corrected series.
The distributions of the deviations are shown in an insert in 
Figure~\ref{fig_acf}. The remaining parameters of the Gaussian fits are: 
The maximum value $N_{\rm max} = 84.8 \pm 2.3$ and  $99.6 \pm 2.4$;
the zero-point shift $\delta_0 = +0.0006 \pm 0.0004$ f.u.\ and 
$+0.0011 \pm 0.0003$ f.u., respectively for the uncorrected and the 
trend-corrected data. We recall that the median error
of the individual data point of the series was estimated at $\sigma = 0.0014$ f.u.

 \placetable{tab:ts}                 % Table 5     delta-series (BTr-4), 3 cols:  un-corr., corr., code

% --------------------------------------- fig4--------------------------------------------------------
% -----------------------------------------------------------------------------------------------------

\subsection{The frequency content}
\label{sec:freq}

The brightening and fading events in the light curve -- relative to its mean level --
appear as up- and down-directed spikes in the  $\delta$ series. 
They tend to occur at time scales shorter than one orbital period of 
$\beta$~Lyr, typically within $0.1P$ to $0.3P$ (Figure~\ref{fig_devs}). 
We counted 27 brightening and 25 fading events during the seven fully
observed cycles of the series, giving the corresponding rates 
$4.0 \pm 0.8$ and $3.6 \pm 0.7$ of such events per orbital period
of $\beta$~Lyr. This is exactly the domain of temporal fluctuations 
most difficult to characterize using ground-based data 
for the orbital period of $\beta$~Lyr. 
Although scales shorter than one orbital period seem to dominate,
we note that variability has components which 
include small multiples of the binary orbital period. Activity at a given phase
may take the form of either a decrease or increase in brightness. For example, 
a series of {\it brightening events\/} took place just before $\phi = 4 - 6$,
while a more conspicuous {\it dip\/} repeated just before the 
very center of the primary eclipse, at $\phi = 7 - 9$ (the approximate
visual estimates gave  spikes occurring at  phases 2.93, 3.95:, 4.93:, 
while  dips were at 5.92:, 6.96, 7.96, 8.96, 9.96:; 
the colon signifies a larger uncertainty).

We used the Auto-Correlation Function (ACF) and the Fourier Transform in its Fast (FFT)
and Discrete (DFT) realizations for the presence of coherent variability in the $\delta$-series.
The orbital phase of $\beta$~Lyr served as the time variable without a correction for
the progressing orbital period change of 5 seconds which took place
within the span of our observations (Section~\ref{sec:ph}).
We considered this effect as unimportant due to the
moderate duration of the run (7 orbital cycles or 1/4 year, in view of the period
change of 19 seconds per year), and the data sampling of one point every 98.3 minutes. 

The ACF of the $\delta$-series in Figure~\ref{fig_acf} 
shows positive correlation at the delay of one orbital period, $P$,
and then consecutive negative spikes at multiples of $P$, starting from the 
delay of $3 \times P$. Thus, whatever pattern emerges in
the deviations, it tends to last for no more than one orbital period, 
but is likely to re-appear after a delay of more than two orbital periods 
in the form of an opposite deviation. 
Obviously, the limited length of the time series 
makes the results for lags larger than about a half
of the length of the data series very uncertain. 
 
% --------------------------------------- fig5--------------------------------------------------------
\begin{figure}[ht]
\begin{center}
\includegraphics[width=0.75\textwidth]{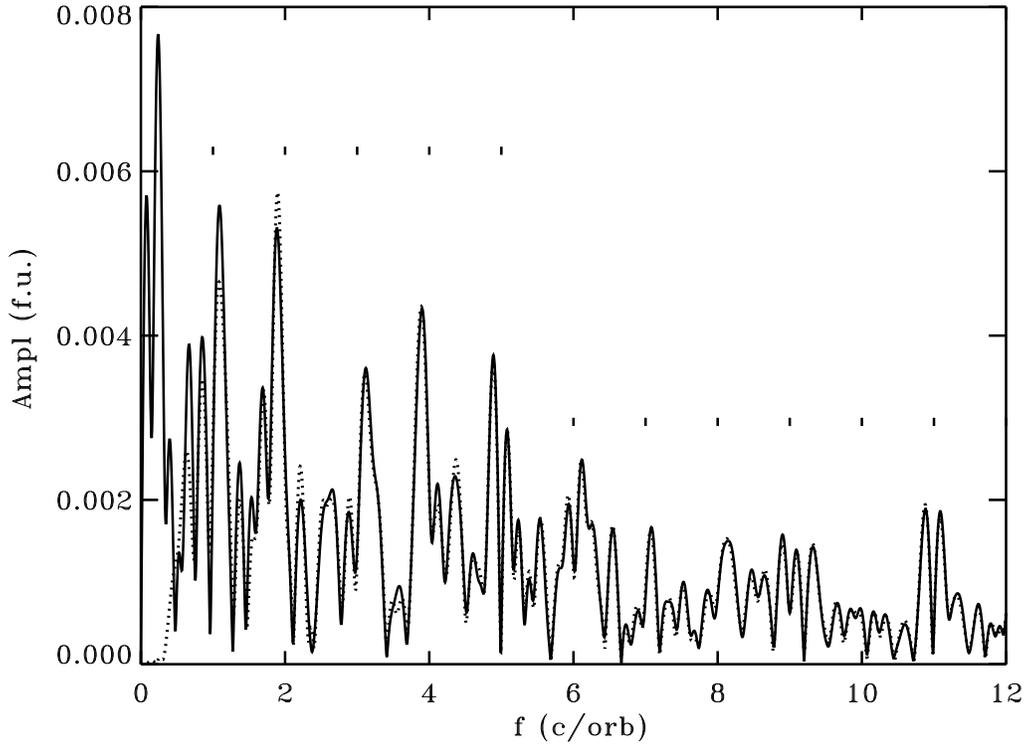}    
\caption{
The Discrete Fourier Transform for the the $\delta$-series,
expressed as cycles per orbital period of $\beta$~Lyr (c/orb). 
The short dashes within the figure mark integer values of the
orbital frequency.
The vertical axis of the figure gives the amplitude of oscillations in flux units
relative to the maximum light of the binary. 
The results shown by the dotted line correspond to the trend-corrected
version of the series.}
\label{fig_DFT}
\end{center}
\end{figure}
% -----------------------------------------------------------------------------------------------------

% frequency spacing 0.13723 c/orb
The FFT and DFT transforms of the $\delta$-series gave identical results.  
Figure~\ref{fig_DFT} shows the low-frequency end of the DFT ($f < 12$ c/orb). 
With the sampling of 98.3 minutes ($0.0052726\,P$) there are 189.6 data points
per orbital period of  $\beta$~Lyr leading to the nominal resolution of
0.1372 c/orb. 
%Since this sampling rate is not  commensurate 
%with the orbital period of the binary, concentration of power per
%frequency bin is poorer for the low frequencies and progressively improves 
%for higher frequencies. 

The uncorrected and the trend-corrected versions of the series give practically
identical results for frequencies above one cycle per binary orbit, 
where well-defined, coherent variability components with frequencies 
corresponding to one and two cycles per period $P$ are easily detectable in the data.
Several higher multiples are also present, but they decrease in size for
higher frequencies and appear to be entirely absent for $f>\!12$ c/orb. 
As expected, the low frequency content is very different in the two versions 
of the series with a very strong drop in amplitudes for $f < 1$ c/orb for
the trend-corrected series; there are no detectable components for $f<0.3$ c/orb.
The strongest frequency in the trend-corrected series is located  
close to 2 c/orb ($1.90\pm0.13$ c/orb) and is due to the two similar eclipses
during one orbital period. Its amplitude is  0.0058 f.u.\ which is only
3.5 times larger than the median data error. Its removal from the FFT and 
re-transformation back to the time series does not change the series in an obvious 
way leaving the general random appearance of the series practically unchanged. 
For the uncorrected series, the largest peak in the transform is located at 0.240 c/orb, 
with the amplitude of 0.0077 f.u.

An estimate of errors in the FFT/DFT transforms is used for normalization of 
the results of the wavelet analysis in Section~\ref{sec:WL}. The assumed error
of the white-noise power was estimated from high frequencies of the transform, 
$f>\!60$ c/orb, at $\sigma_{\rm FFT} \simeq 1.5 \times 10^{-4}$ f.u. 
This estimate is about 4 times larger 
than the expected value using the median of the individual errors, $\sigma = 0.0014$:  
$\sigma_{\rm FFT} = \sigma/\sqrt(1383) =  3.8 \times 10^{-5}$ f.u., 
reflecting the presence of individual, much poorer measurements in the series
(Section~\ref{sec:sat}); however, a contribution of very rapid variability is also not excluded. 

% --------------------------------------- fig6--------------------------------------------------------
\begin{figure}[ht]
\begin{center}
\includegraphics[width=0.35\textwidth,angle=90]{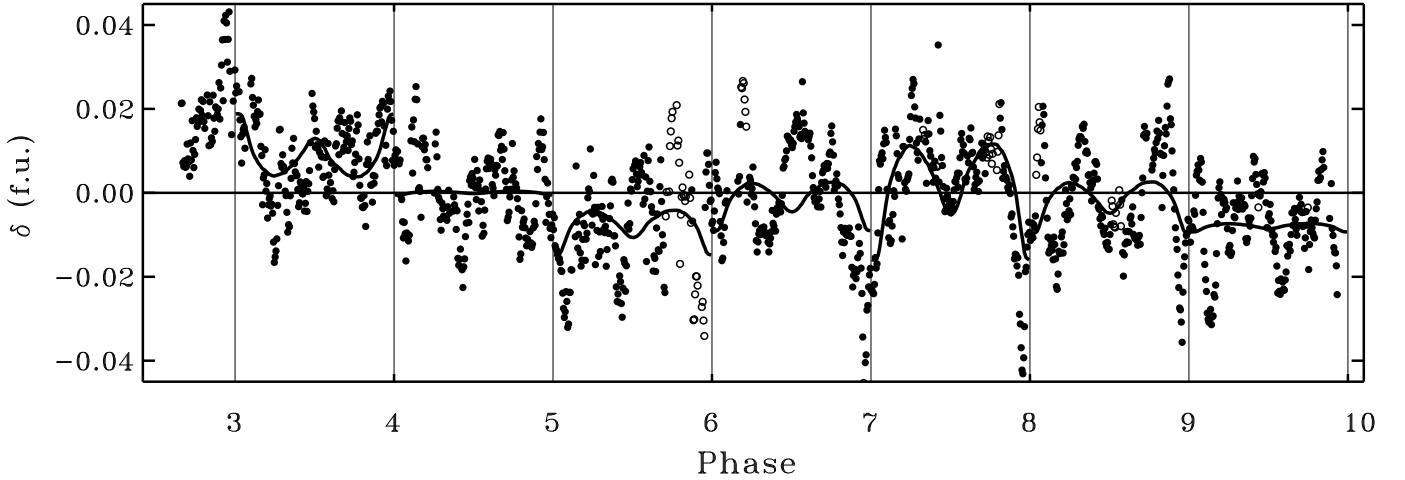}    
\caption{
The deviations of the $\delta$-series without the trend correction, 
fitted by the scaled version of the $\beta$~Lyr mean light curve 
in intervals of full orbital cycles. The gaps in the BTr coverage, 
filled by the UBr data are marked by open circles. 
}
\label{fig_fits}
\end{center}
\end{figure}
% -----------------------------------------------------------------------------------------------------

% --------------------------------------- fig7--------------------------------------------------------
\begin{figure}[ht]
\begin{center}
\includegraphics[width=0.70\textwidth]{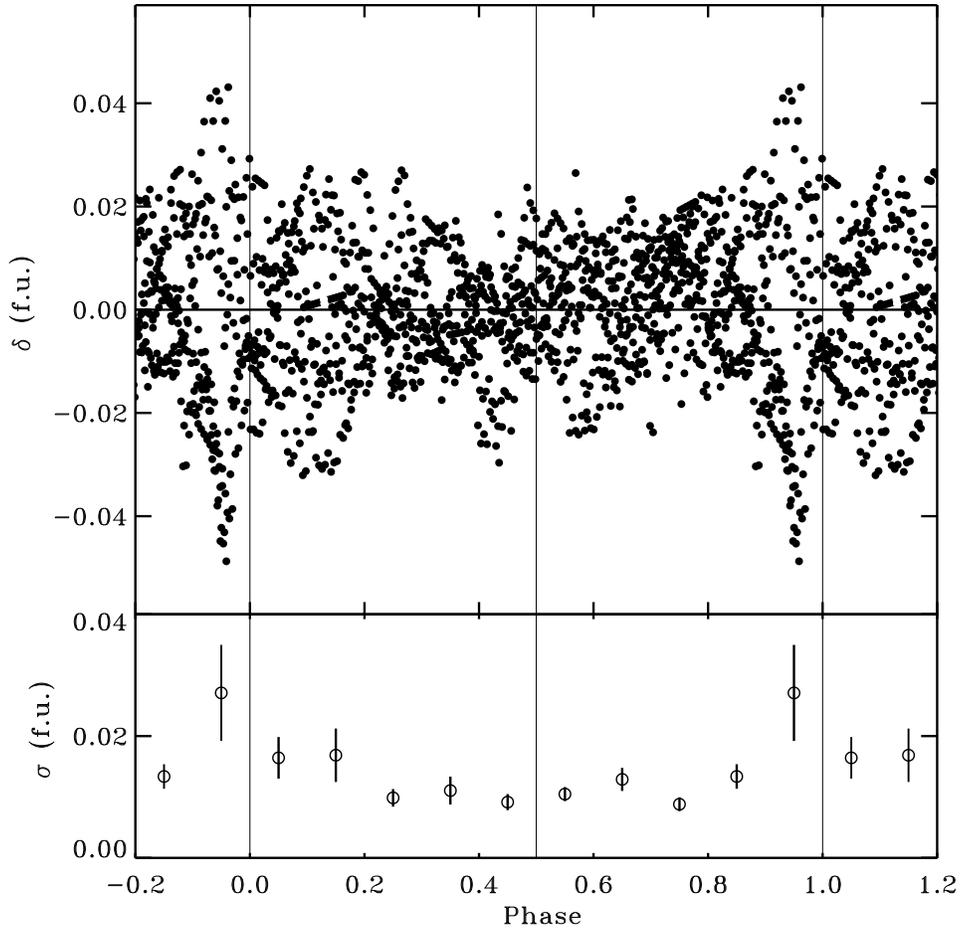}    
\caption{
The $\delta$ deviations of the uncorrected series are shown versus
the fractional phase of the $\beta$~Lyr orbit. The lower panel
gives the standard deviation $\sigma$ computed using  
Gaussian-distribution fits to the deviations in 0.1-wide phase intervals.
}
\label{fig_scatt}
\end{center}
\end{figure}
% -----------------------------------------------------------------------------------------------------

\subsection{The binary-phase dependence}
\label{sec:loc}

Since the orbital period of $\beta$~Lyr together with its harmonics
leaves an imprint in the $\delta$-series, an attempt was made to fit 
the scaled light curve of the binary to individual segments of the
series. This would verify if the deviations are perhaps simply reflections of 
the light-curve shape projecting into the series. An assumption was made of the
exact phase coherence of the $\delta$-deviations with 
the $\beta$~Lyr light curve over each separate orbit interval. Obviously, 
this is highly simplistic as disturbances may emerge or disappear 
at any orbital phase and do not have to last exactly one orbital period. 
No other assumption was made, i.e.\ the light-curve imprint into the
deviations could be either positive or negative corresponding 
to flares or fading dips. The fits are shown in Figure~\ref{fig_fits}. 
In most cases, the fits are inadequate, with the only exception at
cycles $\phi = 7 - 8$ where the deviations do mimic the light curve. 
There is no agreement for earlier cycles, with an inverted dependence 
for the cycle $\phi = 3 - 4$ and no dependence in cycles
$\phi = 4 - 5$ and $9-10$. Thus, we do  see a certain persistence of 
the disturbances with the orbital period of the binary, but generally the
correlation with the light curve is poor. 

An inspection of Figure~\ref{fig_fits} suggests that large, positive and
negative deviations tend to occur at phases just preceding the center of
the primary eclipse when the B6-8~II star is eclipsed by the torus around
the more massive component. The distribution of deviations binned into the 
fractional phase intervals of the $\beta$~Lyr orbit, as in Figure~\ref{fig_scatt}, 
shows the largest spread in the deviations in the orbital phases 
around 0.9 -- 1.0. There, the individual deviations reach $\pm 0.05$ leading
to $\sigma=0.027$, which is 
almost twice what was observed at phases of the secondary eclipse
when the more massive component is eclipsed.  Thus, our observations
point at the region close to the central parts of the torus around the 
more massive component as the location of the photometric disturbances, although
it is of course impossible to tell if this is a permanent feature or a tendency 
which just occurred during the BRITE observations.

% --------------------------------------- fig8--------------------------------------------------------
\begin{figure}[ht]
\begin{center}
\includegraphics[width=0.93\textwidth]{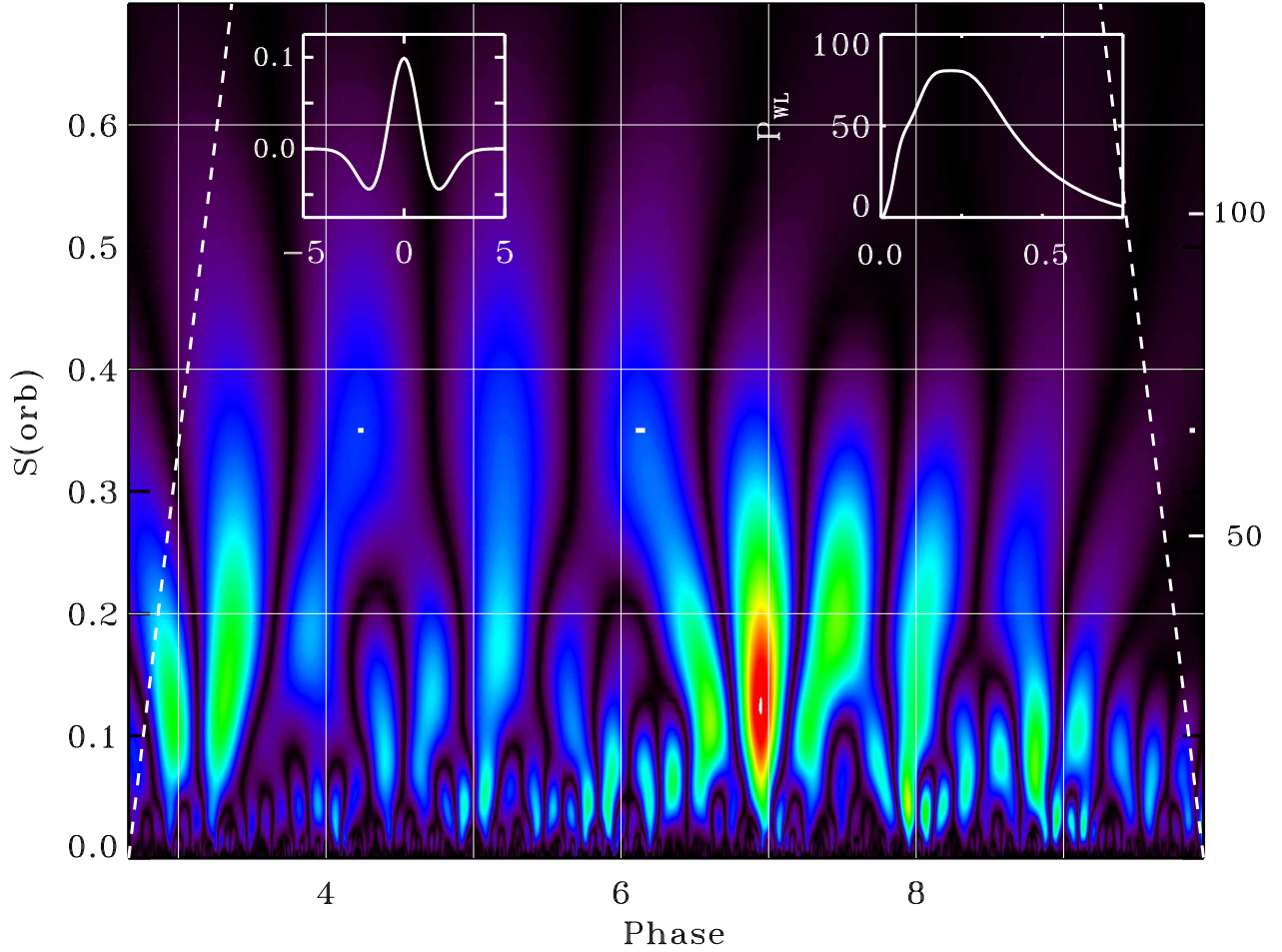}    
\caption{
Wavelet analysis of the $\delta$ series, the version with the low frequency
trend removed (see the text). The horizontal
axis is the orbital phase of $\beta$~Lyr, the same as in previous figures.
The vertical axis is the horizontal scale $S$ of the wavelet analyzing
function -- in this case the DOG-2 or ``Mexican hat'' function -- 
expressed in units of the orbital period, $P$;  the right vertical axis gives 
the scale in numbers of the data points of the series. 
The broken lines delineate areas affected by
edge effects. Short horizontal bars in the middle of the picture give 
phases of points which were not observed and later
interpolated in the equal-step series; 
they do not seem to exhibit any detrimental effects on the results.
The left insert gives the shape of the DOG-2 function; the 
horizontal scale is in units of the data-point spacing while the vertical axis
has been normalized to give the integral of one. 
The right insert shows the variability power 
relative to the white noise of the series with the horizontal scale 
in units of the orbital period of the binary and
the vertical axis giving the variability power normalized by 
the white-noise power, $p_{\rm FFT}$ (see text).
}
\label{fig_WL}
\end{center}
\end{figure}
% -----------------------------------------------------------------------------------------------------

\subsection{Wavelet analysis}
\label{sec:WL}

Wavelet transforms permit characterization of the variability in terms of 
time-scales and of locations of disturbances in the time series. 
In using wavelets, we followed the formulation of \citet{TC1998}, 
where the literature of the subject and  useful recommendations 
are given. Three types of wavelets, Morlet-6, Paul-4 and DOG-2,
were used to perform the continuous 
wavelet transforms of the $\delta$ series: The Morlet-6 wavelet 
detects localized sinusoidal wave packets, while
Paul-4 and DOG-2 are sensitive to localized groups of deviations for which
the  time-scale characterizes the duration of the event.
The numerical descriptors in the wavelet names
provide additional information about the type of the analyzing function,
e.g.\ for Morlet-6 the Gaussian-bound wave packet contains 6 oscillations, 
while DOG-2 is an abbreviation for the second 
derivative of the Gaussian (this function is also known as 
the ``Mexican hat'' wavelet).

Although the Morlet-6 transform has been known to function very well
for detection and frequency characterization of localized and/or evolving wave trains, 
it did not lead to a detection of such periodic events in the $\delta$ series.
We found that both Paul-4 and DOG-2 performed similarly for characterization 
of the spiky disturbances in the series. Since the Paul-4 transform, 
when expressed as the variability power (or square of the amplitude)
and interpreted in terms of the size and time-scale 
gave very similar results to the more popular DOG-2 wavelet, 
we describe here the results only for the latter transform.  
We analyzed both of the $\delta$ series, with and without the low-frequency
trend removed. The uncorrected version was very strongly  
dominated by time scales longer than about $0.5\,P$ due to the large
content of variability power at low frequencies, at time scales $\gg\!P$. 
This forced us to consider only the trend-corrected series, recognizing that this 
will give information only about shorter time scales. 

A wavelet transform unfolds the one-dimensional time series
into a 2-D picture where the abscissa is the time (or in our case the full
$\beta$~Lyr orbital phase), while the ordinate is the scale
of the disturbance, $S$, as shown in Figure~\ref{fig_WL}. 
The scale corresponds to the expansion or
stretching of the assumed wavelet function, in our case the DOG-2. 
As discussed in  \citet{TC1998}, the scales can be related to
the frequencies in the FFT picture; it is also possible
to link the FFT oscillations power to the wavelet-estimated power of the
disturbances, irrespectively of their shape, as long as the analyzing
wavelet function obeys a certain number of conditions. For the figure,
we used the wavelet power normalized to the white-noise power estimated 
from the FFT transform, $\sigma_{\rm FFT} \simeq 1.5 \times 10^{-4}$ f.u.
(Section~\ref{sec:freq}), following the recommendations of \citet{TC1998}.

The variability power over the duration of the run,  $p_{\rm WL}$,
is the sum of the wavelet components 
for the same scales $S$,  as projected into the vertical axis
of Figure~\ref{fig_WL} (the right insert). We see that although most disturbances 
appear with short time scales of  $0.05 P - 0.1 P$ (or 10 -- 20 data points),
in terms of the variability power, the maximum is located at scales of 
$0.2 P- 0.3 P$ (or 40 -- 55 data points); this may signify dominance of the largest
brightening and dimming events in the power budget. 
The  broad range $0.05 P - 0.3 P$ corresponds to time scales between 
0.65 day and 4 days, i.e.\ exactly in the domain of variability which is particularly 
difficult to study from the ground for  $\beta$~Lyr  because of diurnal interruptions.

\subsection{Modeling $\delta$-series as a stochastic process}         % US spelling
\label{sec:DRW}

The light-curve instabilities in $\beta$~Lyr are most likely caused by
the ongoing mass transfer process and may reflect changes in 
the accretion rate onto the hidden, more massive component. 
Accretion is recognized as the key process in active 
galactic nuclei (AGNs), where it is responsible for both typically huge AGN 
luminosity and its significant ($\sim$10\%) random variability. 
The recent years have seen development of modern statistical tools specifically
targeting such aperiodic variations. The damped random walk (DRW) model 
\citep{2009ApJ...698..895K} is
particularly powerful as it describes the variable signal using only
two parameters: (1)~the signal de-correlation time scale $\tau$
and (2)~the modified amplitude of the stochastic signal $\hat{\sigma}$ or, 
equivalently, $SF_\infty$
(\citealt{2010ApJ...708..927K,2010ApJ...721.1014M,2013ApJ...765..106Z}). 

In the DRW model, the variable signal is  correlated for frequencies $\nu$ higher than
$\tau^{-1}$ and resembles  red noise (the higher the frequency, the stronger the correlation), 
with  power spectral density $PSD(\nu)\propto\nu^{-2}$, 
while for frequencies lower than $\tau^{-1}$ it becomes  white noise
($PSD(\nu)\propto\nu^{0} = {\rm const}$). 
Conformity of the time series with the DRW model can be tested by the 
PSD analysis or the structure function (SF) analysis that have 
one additional model parameter, the power-law slope of the correlated noise 
(e.g., \citealt{1984ApJ...284..126S,2016ApJ...826..118K}). 
When analyzed using the SF, which is a measure of the variability amplitude as 
a function of the time difference between points, the DRW process is expected to
show a power-law slope ($\gamma=0.5$) for time scales shorter than $\tau$,
flattening to $\gamma=0$ at time scales longer than $\tau$. 
The signal de-correlation time scale $\tau$ has exactly the same meaning 
in all the three methods: DRW, PSD and SF, so that consistent  results for $\tau$ 
give an additional verification of the model. 
The second parameter of the DRW, the modified amplitude of the stochastic 
process $\hat{\sigma}$, is related to 
$\sigma_\infty$ as $\sigma_\infty=\hat{\sigma}\sqrt{\tau/2}$.
{\bf The data separated in time by more than $\tau$
show variability best described by the white-noise statistics;  $\sigma_\infty$ is 
the amplitude of the white noise, while $SF_\infty$ is the SF amplitude at long time scales,
where $SF_\infty=\sqrt{2}\,\sigma_\infty$. }
All three methods are also interconnected by 
the auto-correlation function (ACF) of the signal, which we generalize here as 
the power exponential (PE; e.g., \citealt{2013ApJ...765..106Z})
\begin{equation}
{\rm ACF}(\Delta t, \tau, \beta) = e^{-(|\Delta t|/\tau)^\beta},
\label{eq:ACFPE}
\end{equation}
where $0<\beta<2$, $\beta=1$ corresponds to DRW, and $\Delta t$ is the time 
difference between points. The conversions between the correlated part of PSD, 
described as the power-law with the slope $\alpha$, the 
structure function (the slope $\gamma$) and the PE (the slope $\beta$) are:
\begin{equation}
\alpha=-(\beta+1),
\end{equation} 
and since the SF slope $\gamma=\beta/2$, we have:
\begin{equation}
\alpha=-(2\gamma+1),
\end{equation}
(e.g., \citealt{2009ApJ...696.1241B,2017ApJ...847..144K}).

% --------------------------------------- fig9--------------------------------------------------------
\begin{figure}[ht]
\begin{center}
\includegraphics[width=0.70\textwidth]{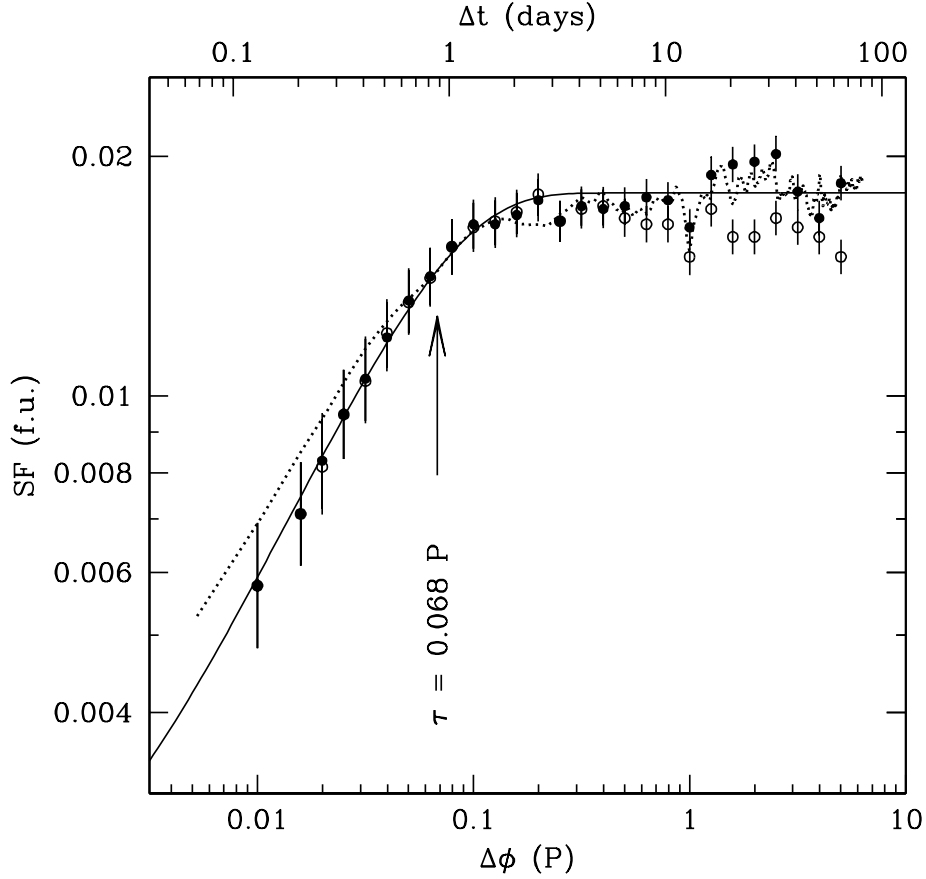}    
\caption{The SF analysis of the $\delta$ series, where the filled circles 
are for the original data while the open circles are for the 
trend-subtracted version of the series. The values of SF have been 
calculated with the IQR method, while the solid line is the best-fit, 
four parameter SF model to the filled dots. 
The de-correlation timescale $\tau$ is marked by an arrow. 
The shallow dips at $\Delta t=1.0 P$ and $4 P$ correspond to an increased correlation, 
that most likely originates from what can be readily seen as repeating 
dips in the $\delta$ series around the fractional phase 0.95, 
as shown in Fig.~\ref{fig_fits}. The dotted line is the ACF
from Section~\ref{sec:freq} converted to SF using Equation~\ref{eq:fullfit} 
with $SF_\infty=0.0179$ f.u.\ and $\sigma_n=0.0014$ f.u.
}
\label{fig_SF}
\end{center}
\end{figure}
% -----------------------------------------------------------------------------------------------------

The SF for the $\delta$-series (Figure~\ref{fig_SF}) was calculated using the inter-quartile 
range (IQR) method (introduced by \citealt{2012ApJ...753..106M}; see Eqns.~(10) 
and (20) in \citealt{2016ApJ...826..118K}). We model it as a 
four-parameter function as in \cite{2016ApJ...826..118K}:
\begin{equation}
SF^2(\Delta t)=SF_\infty^2\left(1-{\rm ACF}(\Delta t, \tau, \beta)\right) + 2\sigma_n^2,
\label{eq:fullfit}
\end{equation}
where the parameters of interest are:
$\tau$, $\beta$, $SF_\infty$, and the photometric noise $\sigma_{n}$.  
The best-fit parameters are: $SF_\infty=0.0179 \pm 0.0003$ f.u.,
%($\sigma_\infty=0.0127\pm 0.0002$ f.u.),
%the PE slope 
$\beta = 1.19 \pm 0.17$, 
%(the SF slope $\gamma=0.60 \pm 0.08$; the PSD slope $\alpha=-2.19 \pm 0.17)$,
$\tau = (0.068 \pm  0.009) P$, $\sigma_n=0.0014$ f.u.\ (fixed),
and the reduced $\chi^2=1.00$.  
These parameters correspond to: $\sigma_\infty=0.0127\pm 0.0002$ f.u., 
the SF slope $\gamma=0.60 \pm 0.08$, the PSD slope $\alpha=-2.19 \pm 0.17$.

We observe two shallow  dips in the flat part of SF 
at $1P$ and $4P$, pointing to a periodic signature in the SF
(see below), otherwise the $\delta$-time series is fully consistent with a stochastic-process 
realization. With the length of $7 P$, the $\delta$-series is much longer than the 
de-correlation timescale ($>$$100$). This enables the usage of an alternative 
method to measure the de-correlation time scale from SF 
introduced recently by \cite{2017ApJ...835..250K}: we find 
$\tau=0.066_{-0.027}^{+0.023}$ $P$ which is consistent with the result above. 

\begin{figure}[ht]
\begin{center}
\includegraphics[width=0.80\textwidth]{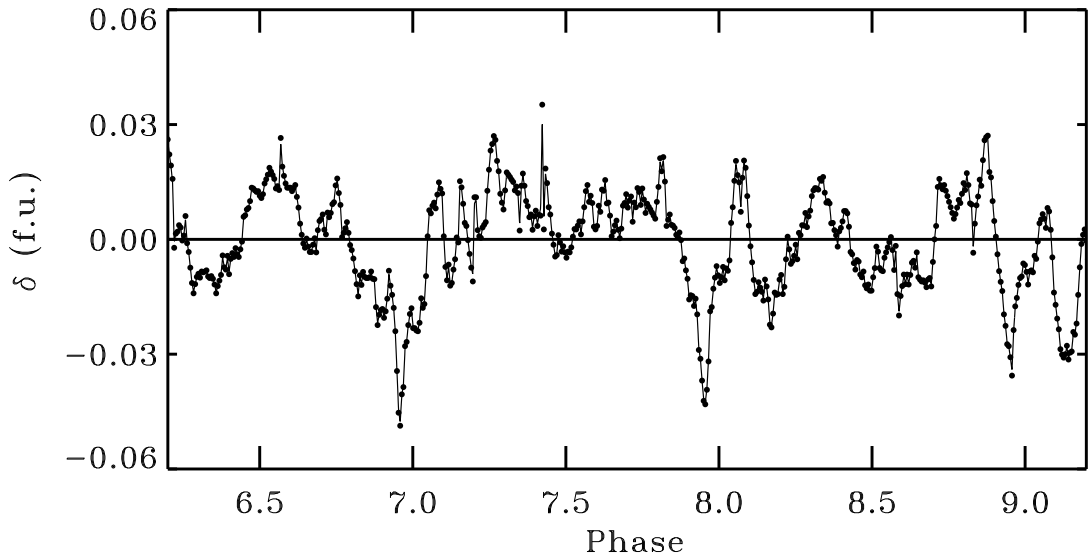}        % expanded version 
\caption{The DRW model to the $\delta$-series is shown by the thin line 
over-plotted on the observed points (dots). The figure covers a part of the $\delta$-series
between phases  $6.2 < \phi < 9.2$  to better show the details of the excellent model fit.
}
\label{fig_DRW}        % same label
\end{center}
\end{figure}

Since the $\delta$-series is much longer than $\tau$, it is possible to 
reliably estimate the DRW parameters (see \cite{2017A&A...597A.128K} 
for a discussion of DRW biases and problems), although the ``red noise'' part 
appears to be slightly steeper ($\gamma=0.60$) 
than that expected for the DRW ($\gamma=0.5$).
The different slope is expected to lead to biased parameters (\citealt{2016MNRAS.459.2787K}), 
with the resulting $\tau$ longer than the true value by about $\sim$40\%. 
We modeled the $\delta$-light curve with the DRW model presented in 
\cite{2010ApJ...708..927K}. The best-fit parameters are: 
$\tau = 0.092_{-0.014}^{+0.021} P$ and $\sigma_\infty=0.0136 \pm 0.004$. 
As expected, because the signal has a stronger correlation than the red noise, 
the de-correlation time scale turned out to be longer than the measurement obtained 
using the SF. 
We present the best DRW model describing the $\delta$-series in Fig.~\ref{fig_DRW}.
In addition to the methods described above, we modeled the PSD 
in the least-squares sense, as described before for the DRW 
% with the DRW PSD 
function (Eq.~(2) in \citealt{2010ApJ...708..927K}), and obtained $\tau=(0.065 \pm 0.014) P$,
again consistent with the above estimates.

% --------------------------------------- fig10  full series modelled -----------------------------------------
%\begin{figure}[ht]
%\begin{center}
%\includegraphics[width=0.27\textwidth,angle=90]{fig_DRW.eps}    
%\caption{The DRW model to the $\delta$-series is shown by the thin line 
%over-plotted on the observed points (dots). 
%}
%\label{fig_DRW}
%\end{center}
%\end{figure}
% -------------------------------------------------------------------------------------------------------------------

% --------------------------------------- fig10 expanded ---------------------------------------------------------
% -----------------------------------------------------------------------------------------------------

Although analysis of the $\delta$-series gives the results perfectly consistent
with the stochastic process assumption, the presence of the weak periodicities with scales 
equal to the orbital period $P$ period and of the associated harmonic frequencies 
(Sections \ref{sec:tser} -- \ref{sec:loc}) requires attention to possibly detrimental 
influence of such coherent signals. In order to check for such influence, 
we analyzed the trend-corrected series (Section~\ref{sec:283}, here called series T1), 
and two additional series obtained by ``brute-force'' removal of the frequencies corresponding
to the scale $P$ (series T2), separately, the scales $P$ and $0.5P$ (series T3).  The removal was
accomplished by setting the FFT frequency components to zero, followed by 
re-transformation back to the time series. While suppression of the lowest frequencies
-- irrespectively if random and/or coherent -- for T1 is a straightforward operation, 
the processes leading to T2 and T3 removed specific coherent signals, possibly 
upsetting the (unknown) relation between the coherent and random components in the series. 

The three test series were subject to the same SF analysis as the original series. 
The results for the trend-corrected series T1 are shown in Figure~\ref{fig_SF} as open circles. 
The low frequencies show a somewhat different slope of the white-noise part of the SF
as expected while the small notch at the delay $\Delta = 1 P$ corresponding
to the dominant coherent signal remains unaltered. 
We do not show the individual SF plots for the series T2 and T3 in order not to 
clutter the figure; basically their random-walk part of SF became more 
irregular and the $\chi^2$ values for Equation~\ref{eq:fullfit} fits noticeably poorer. 
The spread in the determinations of $\tau$ 
reach $\pm 0.018 P$, which we adopt for our best determination,
$\tau = 0.068 P \pm 0.018 P$ or $0.88 \pm 0.24$ day, assuming that the modified,
increased uncertainty adequately -- if possibly rather conservatively -- 
represents the treatment of the coherent signals at low frequencies. 
The time scales within $0.05 P - 0.3 P$ were detected
in the wavelet transform (Section~\ref{sec:WL}), with the shorter scales within this range
dominating in numbers and longer scales dominating in power. The consistency 
of these numbers strongly indicates that this range is the dominant one and that
within this range individual bursts release most of their energy and ``lose identity''
to be replaced by new, uncorrelated ones.

\section{Summary and conclusions}
\label{sec:concl}

The light curve instabilities in $\beta$~Lyr have been known for a long time but 
remained  difficult to characterize in terms of amplitude
and frequency properties, with some previous estimates giving
amplitudes as large as 10\% of the maximum light flux of the star. 
The problems with characterization stemmed from the similarity in time-scales of the
instabilities with diurnal breaks of ground-based observations, 
compounded by difficulties related to standardization of filter photometry 
in the presence of the complex emission-line spectrum of the binary. 
In this work we present an attempt at characterization of the 
photometric instabilities by using a long, nearly continuous time-sequence 
of deviations  $\delta$ from the mean light curve. They 
were determined from the observations of $\beta$~Lyr by two 
red-filter BRITE satellites for over 10 revolutions of the binary. 
The satellite BRITE-Toronto (BTr) provided most of the data
giving uniform flux measurements accurate to 0.0014 f.u. 
(the flux unit is the maximum flux) and sampled at the satellite orbital period ($P$)
of 98.3 minutes for 7.29 binary orbital cycles. 
The data had to be corrected for a newly discovered instrumental problem which 
appears to be caused by radiation damage to the CCD detectors; it
was noted when the more extensive BTr observations were 
compared with the simultaneous (over 4 orbital periods of $\beta$~Lyr) 
observations by the UBr satellite. To define the deviations $\delta$,
we used the mean light curve of $\beta$~Lyr (Section~\ref{sec:lc}) which is very well
determined with the median error per 0.01 phase interval of 0.0036 f.u. 
Although we do not use the mean light curve in this paper, it will be used
in a planned, future investigation \citep[in preparation]{Pav2018}. 

The erratic light variations in $\beta$~Lyr are characterized by a Gaussian distribution 
of the deviations with $\sigma = 0.0130 \pm 0.0004$ f.u.\  
(Sections~\ref{sec:sat} and \ref{sec:283}) with the largest $\delta$ deviations 
not exceeding  $\pm 0.040$ f.u.\ (see the insert in Figure~\ref{fig_acf})\footnote{Although 
variation of the order of 0.01 f.u.\ may seem small,
% bringing perhaps to mind analogy of solar/stellar flares, 
the amount of power released by the accretion phenomena taking place in the
$\beta$~Lyr system is in fact very large in absolute terms. 
Referring to the luminosity of the visible component of $\beta$~Lyr
\citep[Table~1]{Harm2002}, $L/L_\odot  \simeq 5500$, a typical brightening or a dip 
corresponds to a luminosity change amounting to as much as $\simeq 55 L_\odot$.}.
This is less than previously observed, possibly because of the
much more consistent and uniform observational data than ever before, 
but the smaller range of the variations may be related to 
the red-filter bandpass used by the BTr satellite. 
It would be very useful to obtain simultaneous blue and red-band data,
similar to our observations to see if the relatively small
average amplitudes detected here were due to the use of the red bandpass or -- rather -- 
resulted from the high consistency of the  BRITE data. 

The series of the $\delta$ deviations is too short to directly address properties
of the elusive 283 day periodicity noted before (Section~\ref{sec:283}).
However, with the length of 94 days, it is long enough to search 
for periods shorter than one orbital period of $\beta$~Lyr,  
 to frequencies reaching 100 c/orb. 
The Fourier Transform shows several low-frequency harmonics of the 1 cycle per 
orbit, extending to about 5 c/orb or possibly 6 c/orb (Figure~\ref{fig_DFT}), 
but we did not detect the periodicity of of 4.7 -- 4.75 days which had been
suggested by \citet{Harm1996} as linked to the 283 day period. 
However, the 283 day perturbation may have been a part of the slow trend 
with the amplitude of $\pm 0.012$ f.u.\ (Section~\ref{sec:tser}). This
trend is describable by a 5th order polynomial and was fully eliminated,
while its frequency content and amplitude fit very well the picture of the 
Damped Random Walk with red-noise-correlated variations at time 
scales shorter than $\tau = (0.068 \pm 0.018) \, P$ and white noise 
at longer time scales (see below and Section~\ref{sec:DRW}).

The small amplitudes of the coherent, periodic variations  
may be related to the observed sign changes of the dominant 1 c/orb 
perturbation taking place at the same fractional phase: 
We observed strings of positive and negative 
deviations at phases close to about $0.05P$ before the centers of the primary eclipses 
(Figure~\ref{fig_fits}), a tendency confirmed by the Auto-Correlation Function
(Figure~\ref{fig_acf}). 
This led to a larger scatter of the $\delta$ deviations within
the fractional phase range $\phi = 0.9 - 1.0$ (Figure~\ref{fig_scatt}). In contrast,
the  phases of the secondary eclipses show a much smaller orbit-to-orbit scatter. Thus,
although the disturbances seem to be fully random, we noted a directional 
preference for the line of conjunctions with orientations of the highest activity 
when the secondary component and its surrounding torus are in front. 

The wavelet analysis of the $\delta$-series (Section~\ref{sec:WL}, Figure~\ref{fig_WL})
was performed only for the trend-subtracted series with the 5th order polynomial 
removal of scales longer than about $1.2P - 1.35 P$  ($15 - 17$ days), 
otherwise longer scales suppressed the details of the more rapid variations. 
With the DOG-2 or ``Mexican-hat'' analyzing function, instabilities with durations 
within a range of scales around $0.05P - 0.3P$ were the most easily detectable:
The scales at the shorter end of the range, $0.05P-0.1P$, dominated in numbers while longer
scales, within $0.1P-0.3P$,  dominated in terms of the variability power. 
The corresponding scales in time intervals of about 0.65 to 4 days
have been the main difficulty in previous 
attempts in defining mean light curves for $\beta$~Lyr from ground-based observations. 

The dominating time scales were also analyzed using methods developed for 
characterization of erratic variability of AGN and QSO objects. Several statistical tools using
the DRW, SF, and PSD models (Section~\ref{sec:DRW}) clearly show 
the time scale $\tau = (0.068 \pm 0.018) \, P$ or ($0.88\pm0.23)$ day as the location of the break 
where the high-frequency, correlated red-noise changes into white
noise at longer scales, i.e.\ into disturbances independent of each other. The model
describes our observations exceptionally well (Figure~\ref{fig_DRW}) 
confirming the notion of a chaotic accretion process with random bursts 
dissipating their energy in a typical time scale of about $\tau$. 

Since   -- in spite of the vastly different time scales -- 
the stochastic model with the power exponential covariance matrix
of the signal (Eqn.~\ref{eq:ACFPE}) appears to work similarly well for $\beta$~Lyr 
as for  AGNs, we could not resist the temptation to 
relate the $\beta$~Lyr case to what is observed in galactic nuclei. 
\cite{2010ApJ...721.1014M} and \cite{2016ApJ...826..118K} 
report a significant correlation of the optical variability timescale $\tau$
with the black hole mass in AGNs, while $\tau$ appears to be independent of
(or weakly dependent on) the luminosity. An interpretation of this quantity was 
put forward by \cite{2009ApJ...698..895K}, who linked $\tau$ with the orbital or thermal 
time scale in an accretion disk (e.g., \citealt{2006ASPC..360..265C}).
Naively extrapolating the relation from \cite{2016ApJ...826..118K}:
$\log_{10}(\tau/{\rm yr})\propto(0.38\pm0.15)\log_{10}(M/M_\odot)-3.39$, 
by as much as 5--9 dex to the mass of the invisible component of $\beta$ Lyr, 
we find that the expected variability timescale should be $\tau\approx0.4$~d, 
by only a factor of 2 smaller than the actual 
value\footnote{For a full consistency over the huge range of the masses,
the relation should take a particularly simple form: 
$\log_{10}(\tau/{\rm yr})\approx 0.33\log_{10}(M/M_\odot)-3.0$.}.
If indeed the variability in AGNs and $\beta$ Lyr originate in a similar way, 
then the observed timescale of $0.88$~d may be either 
the orbital or the thermal time scale in the accretion disk surrounding 
the more massive companion of $\beta$ Lyr.

\begin{acknowledgements}
Special thanks are due to Dr.\ Petr Harmanec for advice and consultation
on the object that he knows so well. 

The study is based on data collected by the BRITE Constellation satellite mission, 
designed, built, launched, operated and supported by the Austrian Research 
Promotion Agency (FFG), the University of Vienna, the Technical University of Graz, 
the Canadian Space Agency (CSA), the University of Toronto Institute for Aerospace 
Studies (UTIAS), the Foundation for Polish Science and Technology (FNiTP MNiSW), 
and the National Science Centre (NCN). 
The operation of the Polish BRITE satellites is funded by a SPUB grant by the 
Polish Ministry of Science and Higher Education (MNiSW). 

The research of SMR, AFJM and GAW have been supported by the
Natural Sciences and Engineering Research Council of Canada. 
APi acknowledges support from the NCN grant no.~2016/21/B/ST9/01126.
SK acknowledges the financial support of the Polish National Science Center through the
OPUS grant 2014/15/B/ST9/00093 and MAESTRO grant  2014/14/A/ST9/00121.
KP was supported by the Croatian Science Foundation grant 2014-09-8656.
GH acknowledges support by the Polish National Science Center (NCN), 
grant  2015/18/A/ST9/00578.

This research made use of the SIMBAD database, operated at the CDS, 
Strasbourg, France and of NASA's Astrophysics Data System (ADS).
\end{acknowledgements}

% --------------------------------------------------------------------------------------------------------------------
\newpage
% Acta Astronomica \!=\!\!>\! \actaa

%============================================================================
% Table: Setups   Tab.1               \label{tab:setups}

\begin{deluxetable}{lcccc}

\footnotesize
%\tablewidth{0pt}
\tablecaption{Time ranges for the $\beta$~Lyr observations \label{tab:setups}}
\tablenum{1}

\tablehead{\colhead{Setup}  &  \colhead{$t_{\rm start}$} & \colhead{$t_{\rm end}$} & 
      \colhead{$n$}  &\colhead{$N$}
}
\startdata
BTr-2   &  1512.798  & 1520.041 &   2877 &  104  \\
BTr-3   &  1520.101  & 1553.008 & 11682  &  410  \\
BTr-4  &   1553.070  & 1651.967 & 46111  & 1230 \\
BTr-5  &   1653.672  & 1664.800 &   505   &  \nodata  \\     
UBr     &  1590.400   & 1644.874 & 16686 &   479   \\   
BLb    &   1636.345  &  1645.474 &  4919  &   133  \\   
\enddata

\tablecomments{Time:  $t = HJD - 2\,456\,000$.\\
$n$ is the number of individual observations while
$N$ is the number of satellite-orbit averages.\\
The orbital averages were
not used for the BTr-5 setup because of small numbers of 
observations per average data point and large errors. 
}
\end{deluxetable}
% -----------------------------------------------------------------------------------------------------------------------------
% Table: Orbital averages      Tab.2      \label{tab:avg}
% on-line:    tab_asc_avg.txt

\begin{deluxetable}{lccccc}

\footnotesize
\tablecaption{The satellite-orbit average data \label{tab:avg}}
\tablenum{2}

\tablehead{\colhead{$t_{\rm start}$}  &  \colhead{$\phi$} & \colhead{$f$} & 
      \colhead{$\sigma(f)$}  &\colhead{$m$}  &\colhead{Code}
}
\startdata
 1590.4027 & 5.54554 & 0.73552 & 0.00178 & 12 & 1\\
 1590.4725 & 5.55094 & 0.74112 & 0.00262 & 16 & 1\\
 1590.5425 & 5.55634 & 0.74976 & 0.00288 & 13 & 1\\
 1590.6114 & 5.56166 & 0.76306 & 0.00140 &  9 & 1\\
 1590.6856 & 5.56740 & 0.76738 & 0.00418 & 11 & 1\\
 \enddata

\tablecomments{Time $t$, as in the comments to Table~\ref{tab:setups}.\\
The orbital phase of $\beta$~Lyr is computed from \\
$\phi = (t - 1518.6251)/12.943296$ (Section~\ref{sec:ph}). 
The  phase  includes the number of binary orbital cycles from the 
zero epoch. \\
$f$ is the flux in units of the assumed maximum value (see the text)
and $\sigma(f)$ is its error computed from the spread of the contributing 
$m$ individual observations. \\
Code gives the satellite and  setup: \\
1 -- UBr,   2 -- BTr-3,  3 -- BTr-4.  \\
The table is published in its entirety in machine-readable format.
A portion is shown here for guidance regarding its form and content.}

\end{deluxetable}
% -----------------------------------------------------------------------------------------------------------------------------
% Table: Light curve         Tab.3         \label{tab:lc}
% on-line:   tab_asc_lc.txt

% the table widened to reasonable dimensions through addition of phantom columns

\begin{deluxetable}{lcccccc}      % widened artificially

\footnotesize
\tablecaption{The mean light curve of $\beta$~Lyr \label{tab:lc}}
\tablenum{3}

\tablehead{\colhead{$\phi$} & \colhead{~~~~~} & 
                  \colhead{$f$} & \colhead{~~~~~} &  
                  \colhead{$\sigma(f)$}  & \colhead{~~~~~} &
                  \colhead{$n$}
}
\startdata  
 0.00492 & & 0.50679 & & 0.00542  & & 17\\
 0.01455 & & 0.50429 & & 0.00655  & & 15\\
 0.02440 & & 0.50492 & & 0.00557  & & 15\\
 0.03449 & & 0.50961 & & 0.00434  & & 17\\
 0.04514 & & 0.52418 & & 0.00398  & & 15\\
\enddata

\tablecomments{$\phi$ is the mean fractional phase of $\beta$~Lyr,
as in  Comments to Table~\ref{tab:avg}.\\
$f$ is the mean flux calculated per  interval of 0.01 in phase,
 while $\sigma(f)$ is the error computed from the spread of the 
 contributing $n$ individual satellite-orbit points.  \\
The table is published in its entirety in machine-readable format.
A portion is shown here for guidance regarding its form and content.}

\end{deluxetable}
% -----------------------------------------------------------------------------------------------------------------------------
% Table: Statistics of      Tab.4         \label{tab:devs}

\begin{deluxetable}{cccccccc}   

\footnotesize
\tablecaption{The equal-step $\delta$ series \label{tab:devs}}
\tablenum{4}

% Series  Start-ph    End-ph     a0        a1       n-obs      N   \%   \\

\tablehead{\colhead{Setup}  &  \colhead{$\phi_1$} & \colhead{$\phi_2$} & 
      \colhead{$a_0$}  &  \colhead{$a_1$}   & \colhead{$N$} & \colhead{$M$} &
      \colhead{\%}
}
\startdata
BTr-3  &  0.114 &  2.656  &  0.1147 &  0.0052726  &  410  &  483  &  84.9  \\
BTr-4  &  2.661 &  10.301 (9.950) &  2.6617 &  0.0052726  & 1230  & 1450 (1383) & 84.8 (88.9)  \\
UBr    &  5.546  &   9.754  &  5.5456 &  0.0053880  &  479   & 781  &  61.3    \\
\enddata

\tablecomments{
$\phi_1$ and $\phi_2$ are the start and end orbital phases of $\beta$~Lyr,
as explained in  Comments to Table~\ref{tab:avg}.
% computed from $\phi = (t - 1518.6251)/12.943296$ and $t =$ HJD$- 2\,456\,000$.
The equal-step phases were calculated using $\phi = a_0 + a_1 \, i$, where $0 \le i \le M-1$.
The last column gives the coverage of the series expressed as the
percentage of the number of actually observed satellite-orbit averages, $N$,
relative to the length of the equal-step series, $M$.
The BTr-4 series was truncated to $M=1383$ for a detailed analysis of the time series; 
the corresponding numbers of the end phase $\phi_2$ and the percentage of observed 
equal-step intervals are given in parentheses. \\
For the $\delta$ series resulting
from filling the BTr-4 missing data by the UBr observations: $N = 1299$ and $M=1383$, 
giving the coverage 93.9\%.
}
\end{deluxetable}

% ------------------------------------------------------------------------------------------------------------------------------
% Table: The final delta-series     Tab.5       
%  on-line:    tab_asc_ts.txt

\begin{deluxetable}{crrcc}                % added fake columns
%\begin{deluxetable}{rrc}       

\footnotesize
\tablecaption{The equal-step $\delta$ series \label{tab:ts}}
\tablenum{5}
 
\tablehead{\colhead{$\delta_1$} & \colhead{~~~~~} & \colhead{$\delta_2$} & \colhead{~~~~~} & \colhead{Code}
}
\startdata
   0.02128 &  &  $0.00570$  &  & 1  \\
   0.02138 &  &  $0.00581$  &  & 1  \\
   0.00719 &  &  $-0.00839$ &  & 1  \\
   0.00627 &  &  $-0.00931$ &  & 1  \\
   0.00767 &  &  $-0.00790$ &  & 1  \\
\enddata

\tablecomments{
% rewrite
The columns $\delta_1$ and $\delta_2$ give the deviations expressed in the flux 
units for the un-corrected and trend-corrected series
(Section~\ref{sec:283}), while Code signifies: \\
0 -- missing point, interpolated using entries with Codes 1 and 2, \\
1 -- observed, BTr-4 setup,\\
2 -- observed by UBr, interpolated into the equal-step BTr-4 series.\\
The phases can be restored using the values of $a_0$ and $a_1$ for
the BTr-4 entry in Table~\ref{tab:devs} with $M = 1383$. \\
The table is published in its entirety in  machine-readable format.
A portion is shown here for guidance regarding its form and content.}

\end{deluxetable}
% ------------------------------------------------------------------------------------------------------------------------------

\end{document}